\documentclass[conference]{IEEEtran}
\IEEEoverridecommandlockouts

\usepackage[
  backend=biber,
  style=ieee,
  minnames=1,
  maxcitenames=2, maxbibnames=6
]{biblatex}
\usepackage{amsmath,amssymb,amsfonts}
\usepackage{algorithmic}
\usepackage{graphicx}
\usepackage{textcomp}
\usepackage{xcolor}
\usepackage{hyperref}
\usepackage{booktabs}
\def\BibTeX{{\rm B\kern-.05em{\sc i\kern-.025em b}\kern-.08em
    T\kern-.1667em\lower.7ex\hbox{E}\kern-.125emX}}
\begin{document}

\title{Anomaly Detection in IEC-61850 GOOSE Networks: Evaluating Unsupervised and Temporal Learning for Real-Time Intrusion Detection}

\author{
\IEEEauthorblockN{Joseph Moore}
\IEEEauthorblockA{\textit{Computer Science} \\
\textit{Boise State University}\\
Boise, Idaho USA}
}

\maketitle

\begin{abstract}
The IEC-61850 GOOSE protocol underpins time-critical communication in modern digital substations but lacks native security mechanisms, leaving it vulnerable to replay, masquerade, and data injection attacks. Intrusion detection in this setting is challenging due to strict latency constraints (sub-4~ms) and limited availability of labeled attack data. This paper evaluates whether unsupervised temporal modeling can provide effective and deployable anomaly detection for GOOSE networks. Five models are compared on the ERENO IEC-61850 dataset: a supervised Random Forest baseline, a feedforward Autoencoder, and three recurrent sequence autoencoders (RNN, LSTM, and GRU). The supervised Random Forest achieves the highest detection performance (F1\,=\,0.9516) but fails to meet real-time constraints at 21.8~ms per prediction. All four unsupervised models satisfy the 4~ms requirement, with the GRU achieving the best accuracy to latency tradeoff among them (F1\,=\,0.8737 at 1.118~ms). A cross-environment evaluation on an independent dataset shows that all models degrade under distribution shift. However, recurrent models retain substantially higher relative performance than the supervised baseline, suggesting that temporal sequence modeling generalizes better than fitting labeled attack distributions. Anomaly thresholds for the unsupervised models are selected on a held out validation partition to avoid test set leakage. These results support unsupervised temporal models as a practical choice for real-time GOOSE intrusion detection, particularly in environments where labeled training data may be unavailable or where large-scale deployment across diverse substations is required.
\end{abstract}

\section{Introduction}
The modernization of electrical power infrastructure has introduced digital substations governed by the IEC-61850 standard, enabling interoperability between intelligent electronic devices (IEDs) through a common communication framework. A cornerstone of this framework is the Generic Object-Oriented Substation Event (GOOSE) protocol, which delivers time-critical protection and control signals directly over the Ethernet data link layer. By bypassing both the network and transport layers, GOOSE achieves the sub-4~ms latency required by the standard for reliable fault isolation and power system stability \cite{IEC61850_GOOSE}.

However, this architectural decision comes at a significant security cost. The GOOSE mapping defined in IEC-61850-8-1 includes no native encryption, authentication, or integrity verification mechanisms \cite{IEC61850_GOOSE}. As substations transition from isolated serial networks to IP-connected Ethernet environments, this design choice exposes critical infrastructure to a range of cyberattacks. Adversaries can inject fabricated GOOSE messages to trigger false circuit breaker operations, replay legitimate historical messages to cause inappropriate switching events, or masquerade as authorized IEDs to manipulate substation state. The consequences of a successful attack can range from localized equipment damage to large-scale power outages.

Intrusion detection systems designed for IT environments are poorly suited to substation deployments. IEC-61850 traffic exhibits highly structured, deterministic behavior that differs fundamentally from general enterprise network traffic, and the strict real-time requirements of the substation environment impose hard constraints on detection latency. Furthermore, the extreme class imbalance between normal and attack traffic, combined with the need to detect novel attack variants not seen during training, limits the effectiveness of standard supervised classifiers \cite{lozano2026explainable}.

This paper investigates whether unsupervised learning approaches, and temporal modeling in particular, offer practical advantages for GOOSE intrusion detection. The following contributions are made:

\begin{figure*}[h!]
    \centering
    \includegraphics[width=1.0\textwidth]{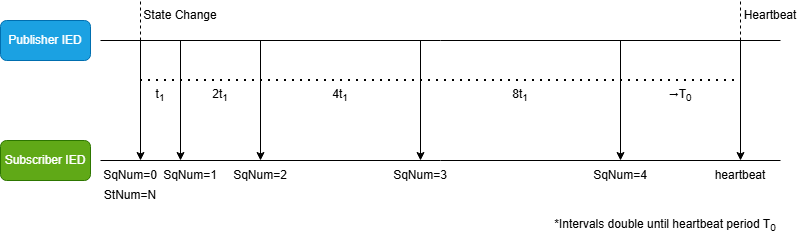}
    \caption{GOOSE Retransmission Sequence}
    \label{fig:retransmission}
\end{figure*}

\begin{itemize}
    \item Implementing and evaluating five detection models on the ERENO IEC-61850 dataset: a supervised Random Forest, an unsupervised feedforward Autoencoder, and three unsupervised recurrent sequence autoencoders (LSTM, RNN, and GRU).
    \item Evaluating all models not only on detection accuracy but also on inference latency, directly addressing the real-time deployment constraint imposed by the 4~ms GOOSE protection requirement.
    \item Demonstrating that all recurrent temporal models satisfy the 4~ms latency budget while also achieving competitive F1-scores, with the GRU achieving the best F1-score (0.8737) among the unsupervised models.
    \item Selecting anomaly detection thresholds on a held out validation partition rather than the test set, ensuring that reported metrics are not inflated by threshold leakage.
    \item Analyzing the extent to which sequential GOOSE message modeling captures temporal attack signatures that the static feedforward Autoencoder misses.
    \item Evaluating cross-environment generalizability by testing all trained models on a second IEC-61850 dataset (from the work of de Oliveira et al.~\cite{DEOLIVEIRA2025104197}) from a different physical testbed without retraining, revealing that the supervised Random Forest generalizes substantially more poorly than the unsupervised temporal models.
\end{itemize}

The remainder of this paper is structured as follows. Section~\ref{sec:background} reviews relevant literature and provides background on the IEC-61850 GOOSE protocol and its security properties. Section~\ref{sec:methods} describes the dataset and methodology. Section~\ref{sec:results} presents experimental results. Section~\ref{sec:discuss} provides a discussion of the findings. Section~\ref{sec:generalize} evaluates cross-environment generalizability on a second dataset, and Section~\ref{sec:conclusion} concludes and discusses future work.

\section{Background and Related Work} \label{sec:background}

Research into IEC-61850 security has grown substantially as digital substation deployments have expanded. This section surveys key works addressing GOOSE protocol vulnerabilities, dataset generation, and machine learning based detection approaches.

\subsection{IEC-61850 and the GOOSE Protocol}
IEC-61850 is the international standard for communication networks and systems in substations. It defines an abstract communication service interface that is mapped to concrete protocols through specific communication service mappings. There are two primary communication methods in the IEC-61850 standard. For continuous analog data monitoring, a stream of Sampled Value (SV) transmissions tracks the current network state. For time critical events such as circuit breaker trips, protection relay signals, and interlocking commands, the GOOSE mapping defined in IEC-61850-8-1 is used.

GOOSE operates on a publisher-subscriber model directly at the Ethernet data link layer (EtherType 0x88B8), bypassing the network (IP) and transport (TCP) layers entirely. This architectural choice is deliberate since eliminating TCP/IP handshaking and retransmission overhead is what makes sub-4~ms delivery reliably achievable. Each GOOSE frame encodes the current state of a monitored data object on the substation network. For example, whether a circuit breaker is open or closed along with metadata fields used by subscribers to reason about message validity and ordering.

As shown in Figure~\ref{fig:retransmission}, when a monitored value changes state, the publishing IED immediately transmits a GOOSE frame and then retransmits it at exponentially increasing intervals until the heartbeat period is reached. Each frame carries a Status Number ($StNum$), which increments with each new state change, and a Sequence Number ($SqNum$), which increments with each retransmission of the same state. For instance, a state change at $t_1$ ($StNum=3, SqNum=0$) is followed by retransmissions at $t_2$ ($SqNum=1$), $t_4$ ($SqNum=2$), and so forth until the heartbeat interval is reached. At that point, $SqNum$ resets to 0 and $StNum$ increments to signal a new state or a keep-alive message. Monitoring devices utilize the strict correlation between these counters and arrival times to identify dropped packets and distinguish legitimate state transitions from replayed traffic.

This retransmission mechanism, combined with the absence of transport layer overhead, enables GOOSE to meet the 4~ms trip time requirement for distance protection. However, the protocol was originally designed for isolated networks where physical access control was assumed to provide sufficient security. As substations have migrated to larger Ethernet connected environments, this assumption no longer holds. Any device on the local area network segment can receive, replay, or fabricate GOOSE frames without authentication.

\begin{figure*}[h!]
    \centering
    \includegraphics[width=0.85\textwidth]{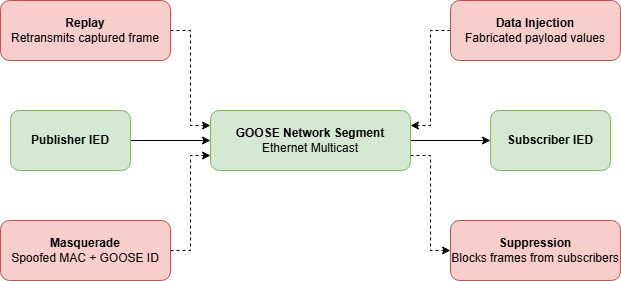}
    \caption{Attack classes during GOOSE transmission}
    \label{fig:attacks}
\end{figure*}

\subsection{Attack Taxonomy}
The principal attack classes relevant to GOOSE networks are as follows. In a \emph{Replay} attack, a legitimate captured GOOSE frame is retransmitted at a later time to cause a spurious state change. Anomalous $StNum$ and $SqNum$ progressions relative to timing are the primary indicators. In a \emph{Data Injection} attack, a fabricated GOOSE frame with manipulated payload values (such as a false circuit breaker status) is inserted into the network. \emph{Masquerade} attacks involve an adversary impersonating a legitimate IED by spoofing its source MAC address and GOOSE identifiers. \emph{Message Suppression} attacks prevent legitimate frames from reaching subscribers, either through physical interference or by flooding the network \cite{s21041554}. High rate flooding can also arise from legitimate cascading events that trigger simultaneous GOOSE transmissions across the network, a condition commonly referred to as a GOOSE storm~\cite{kabirquerrec:hal-01237725}.

\subsection{The ERENO Dataset}

A persistent challenge in IEC-61850 IDS research has been the absence of high fidelity, publicly available labeled datasets. Quincozes et al.\ \cite{ERENO2024} address this gap with ERENO, an open-source framework that integrates network traffic simulation with physical power system modeling. ERENO generates GOOSE and SV traffic across eight scenarios (one normal operational class and seven attack classes including Data Injection, Replay, and Masquerade) modeled on a real electrical transmission line. Crucially, ERENO also simulates legitimate physical faults such as short circuits, which is essential for reducing false positives in realistic deployments. The authors demonstrate that even simple models such as the J48 decision tree achieve high detection performance on the generated features, validating the dataset's representativeness \cite{ERENO2024}.

The ERENO IEC-61850 dataset \cite{ERENO2024} used in this work contains 69 features derived from GOOSE and SV protocol fields. These include raw instantaneous and Root Mean Square (RMS) power measurements from busbar and motor-side sensors, GOOSE frame metadata fields, and a set of engineered delta features capturing temporal relationships between consecutive messages: $StNum$ difference (\texttt{stDiff}), $SqNum$ difference (\texttt{sqDiff}), inter-message timing (\texttt{timestampDiff}, \texttt{tDiff}), and time since last state change (\texttt{timeFromLastChange}). The dataset is partitioned into a pre-split training set and test set spanning normal traffic and seven attack scenarios. The framework also supports simulation in varied configurations, enabling evaluation across diverse substation setups.

\subsection{Feature Selection and Lightweight Models}
Quincozes et al.\ \cite{s21041554} address the challenge of deploying IDS on resource-constrained embedded hardware within digital substations. They combine Deep Packet Inspection with a Genetic Algorithm for feature selection, identifying the most informative protocol features before training a Convolutional Neural Network (CNN) and a sparse neural network optimized via Differentiable Architecture Search. The work demonstrates that dimensionality reduction is critical not only for accuracy but for practical deployment, as inference time and CPU/RAM utilization on embedded hardware are primary constraints. This motivates our own feature reduction strategy, which prioritizes temporally informative delta features over high cardinality categorical fields.

\subsection{Feature Relationship Modeling}
Wang et al.\ \cite{wang2025research} observe that existing methods struggle with the diverse data types present in IEC-61850 features (both numeric measurements and categorical protocol fields) and with capturing complex inter-feature relationships. Their ATCV model integrates feature selection with Fuzzy Triadic Concept Analysis to construct multidimensional feature vectors that encode relationships between protocol attributes and attack classes. While ATCV achieves strong classification performance on the ERENO dataset, the authors acknowledge that the computational complexity of generating fuzzy triadic concepts may limit real-time applicability, making it better suited as an analysis tool than an inline network detector.

\subsection{Unsupervised and Anomaly-Based Detection}
Jay et al.\ \cite{jay2022unsupervised} provide a direct comparison of two unsupervised techniques (DBSCAN clustering and Autoencoders) for detecting GOOSE payload corruption anomalies. The study demonstrates that Autoencoders trained exclusively on normal traffic can identify anomalies through elevated reconstruction error without requiring labeled attack examples. However, the work focuses narrowly on payload corruption and does not address temporal or timing-based attack vectors, which are central to replay and masquerade scenarios.

Lozano-Paredes et al.\ \cite{lozano2026explainable} extend the unsupervised autoencoder paradigm with an asymmetric architecture that jointly monitors semantic integrity and temporal availability of GOOSE messages. A key contribution is intrinsic explainability: the model identifies the specific protocol feature that triggered each alarm, addressing the black-box criticism common to deep learning IDS. The study achieves a detection rate exceeding 99\% with a false positive rate below 5\% on real-world traffic, though the authors note that temporal feature reliance may introduce false positives during natural network jitter.

\subsection{Recurrent Architectures for Temporal Detection}
De Oliveira et al.\ \cite{DEOLIVEIRA2025104197} provide the most directly relevant prior work, systematically evaluating LSTM, GRU, and their bidirectional variants for GOOSE IDS using a Hardware-in-the-Loop (HIL) simulation environment. The study demonstrates that recurrent architectures significantly outperform static classifiers on event-driven attacks (particularly Masquerade and Replay) by capturing the sequential state transitions inherent in GOOSE communications. The authors note, however, that RNN-based models are sensitive to the chosen sequence window length, raising practical deployment concerns that this paper addresses through explicit inference throughput evaluation.

\section{Methodology} \label{sec:methods}

The evaluation framework is designed to determine which model configurations best satisfy the joint requirements of detection accuracy, inference latency, and cross-environment transferability imposed by the IEC-61850 GOOSE environment.

\subsection{Model Architectures}
Three model families are evaluated to compare supervised learning, static anomaly detection, and temporal anomaly detection.

\textbf{Random Forest Baseline:}
The Random Forest classifier serves as a supervised baseline. It is trained on the full labeled training set with binary targets (normal vs.\ attack) and \texttt{class\_weight=`balanced'} to address the substantial class imbalance present in the dataset. The model is trained using 50 trees and a minimum sample split threshold of 10 with default feature selection parameters. Because the Random Forest operates on individual messages rather than sequences, it provides a useful performance reference against which the unsupervised models can be compared. Latency is measured using scikit-learn's default \texttt{predict} interface with batch size 1. The high per-call overhead of this interface is a known characteristic of the library and likely overstates the architectural latency of decision forest inference. A hardware accelerated or batch optimized deployment could reduce this cost, though it would remain unsuitable for inline IDS without further optimization.

\begin{figure}[h!]
    \centering
    \includegraphics[width=0.36\textwidth]{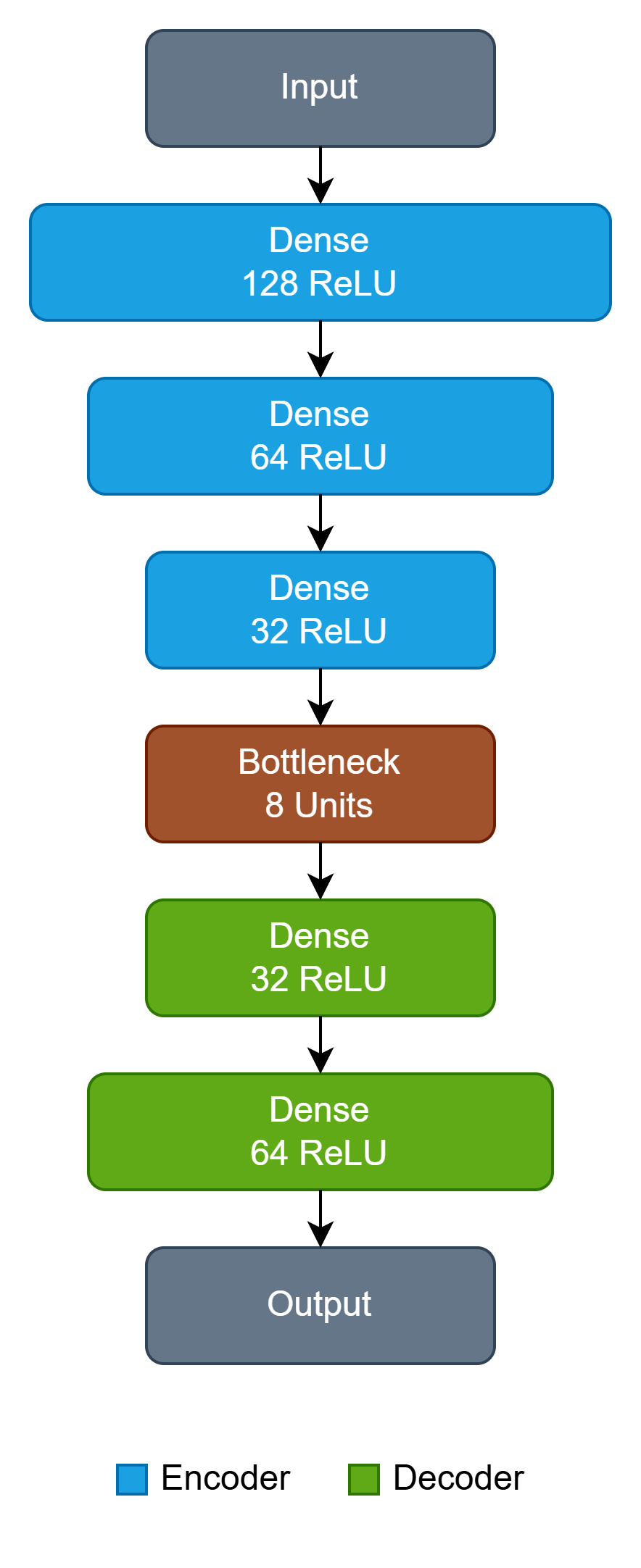}
    \caption{Autoencoder Model Structure}
    \label{fig:autoencoder}
\end{figure}

\textbf{Feedforward Autoencoder:}
The feedforward Autoencoder is trained exclusively on samples from the normal class in order to learn a compact representation of legitimate GOOSE traffic patterns. The encoder network compresses the input feature vector through three fully connected layers of 128, 64, and 32 units using ReLU activations. The decoder mirrors this structure to reconstruct the original input. The model is trained using mean squared reconstruction error as the loss function. The architecture is illustrated in Figure~\ref{fig:autoencoder}.

During inference, reconstruction error is computed for each sample as the mean squared difference between the input and the reconstructed feature vectors. Because the model has only been exposed to normal traffic during training, anomalous inputs are expected to produce higher reconstruction errors. An anomaly detection threshold is applied to these values. The threshold is selected on a held out validation partition (10\% of the training data, withheld before model fitting) by maximizing the Youden Index (TPR$-$FPR) on the validation ROC curve. This avoids inflating test set metrics through threshold leakage.

\begin{figure}[h!]
    \centering
    \includegraphics[width=0.25\textwidth]{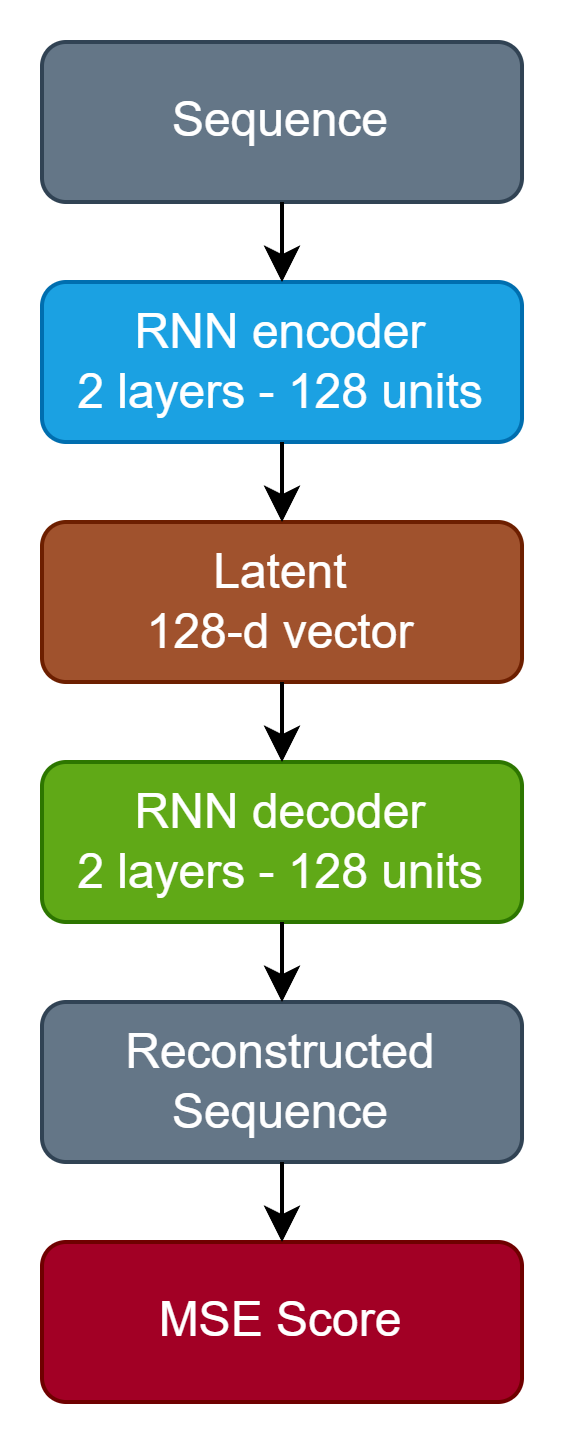}
    \caption{RNN Models Structure}
    \label{fig:RNN}
\end{figure}

\textbf{RNN Autoencoders:}
The Recurrent Neural Network (RNN) Autoencoders extend the feedforward architecture to capture temporal dependencies between consecutive GOOSE messages. Input features are organized into sliding windows of $SEQ\_LEN$ consecutive messages, forming tensors of dimension $SEQ\_LEN \times N$, where $N$ is the number of retained features after preprocessing. The same basic encoder--decoder structure (illustrated in Figure~\ref{fig:RNN}) is used for the LSTM, plain RNN, and GRU variants.

The encoder processes the input sequence and produces a compressed latent representation of temporal behavior within the window. The decoder reconstructs the sequence from this representation. As with the feedforward Autoencoder, training uses only normal traffic windows so that the model learns the characteristic temporal patterns of legitimate GOOSE communication, including expected $StNum$/$SqNum$ progressions and inter-message timing. The anomaly threshold is likewise selected on the held out validation partition.

At inference time, reconstruction error is computed across all time steps and features within each window. A window is labeled as an attack if any message within the sequence belongs to an attack class.

\subsection{Feature Sets}
Three feature presets are used in hyperparameter tuning. The \emph{GOOSE+SV} preset retains all GOOSE and Sampled Values features (21 features). The \emph{GOOSE-only} preset retains GOOSE protocol fields without SV measurements. The \emph{de Oliveira} preset uses the reduced 15-feature set recommended by de Oliveira et al.~\cite{DEOLIVEIRA2025104197}, which prioritizes GOOSE temporal delta features (\texttt{stDiff}, \texttt{sqDiff}, \texttt{timestampDiff}, \texttt{tDiff}, \texttt{timeFromLastChange}) and omits high cardinality categorical fields. For all presets, the absolute timestamp column is excluded as it encodes temporal position rather than temporal dynamics and would not generalize across deployments.

Each model is tuned independently across presets and the best performing preset per model is reported in Section~\ref{sec:results}. The Random Forest is evaluated on the GOOSE+SV preset (highest feature coverage). The LSTM is evaluated on GOOSE+SV while the GRU and RNN are evaluated on the de Oliveira preset. While interpreting cross-model F1 comparisons one should keep in mind that differences in reported performance may partially reflect feature set differences in addition to architectural differences. To isolate architectural effects, a standardized comparison on a single shared feature set is recommended in future work.

\subsection{Evaluation Metrics}

All models are evaluated using Accuracy, Precision, Recall, F1-Score, and Area Under the Receiver Operating Characteristic Curve (AUC-ROC). These metrics provide complementary perspectives on detection performance given the significant class imbalance in the dataset.

In addition to classification performance, inference latency is measured for each model to assess deployment feasibility. Latency is defined as the time passed per prediction during inference on the test dataset using a batch size of one, approximating sequential message arrival in a real-time deployment. Because protection signals in IEC-61850 networks require response times on the order of 4~ms, models must operate at sufficiently low latency to avoid introducing unacceptable delays.

For each RNN-based Autoencoder, additional experiments evaluate the impact of sequence window length ($SEQ\_LEN \in \{5, 10, 20, 50\}$) on both detection performance and computational overhead.

\section{Experiments} \label{sec:results}

\subsection{Data Preprocessing}
The raw CSV files in the public ERENO dataset contain a malformed header row (a spurious leading ID column absent from the data rows), which is corrected by discarding the CSV header and applying a manually verified column name mapping. Columns with no values are dropped, and shared columns between the train and test splits are retained. Numeric features are standardized using a StandardScaler fitted on the training set. Categorical features (including MAC addresses, EtherType, protocol identifiers, GOOSE reference strings, and boolean flags) are one-hot encoded using a vocabulary fitted on training data, with unseen test values handled via an ignore policy.

A held out validation partition (10\% of the training data, stratified by class label) is withheld before any model fitting. This partition is used exclusively for anomaly threshold selection in the unsupervised models and is never used during training or reported test evaluation.

High cardinality categorical columns that contribute primarily to feature dimensionality rather than predictive signal (including MAC address fields, GOOSE reference strings, and protocol constants) are dropped to reduce memory pressure and improve generalizability across deployments.

Table~\ref{tab:distributions} summarizes the class distribution in the training and test splits. The degree of class imbalance (approximately 93\% normal traffic) directly motivates the use of \texttt{class\_weight=`balanced'} in the Random Forest and the normal-only training strategy in the Autoencoder and RNN models.

\begin{table}[ht]
    \centering
    \caption{Attack Class Distributions}
    \label{tab:distributions}
    \begin{tabular}{|l|l|l|}
        \hline
         Class & Train Distribution & Test Distribution \\
         \hline
         normal                   & 93.36\% & 93.21\% \\
         high\_stNum              & 1.32\%  & 1.31\%  \\
         injection                & 1.32\%  & 1.31\%  \\
         inverse\_replay          & 0.88\%  & 1.02\%  \\
         masquerade\_fake\_fault  & 0.58\%  & 0.58\%  \\
         masquerade\_fake\_normal & 0.59\%  & 0.58\%  \\
         poisoned\_high\_rate     & 0.62\%  & 0.62\%  \\
         random\_replay           & 1.32\%  & 1.31\%  \\
        \hline
    \end{tabular}
\end{table}

\subsection{Detection Performance}

Table~\ref{tab:results} reports Accuracy, Precision, Recall, F1-Score, and AUC-ROC for all five models using their best tuned hyperparameter configurations. The feature preset used for each model is noted in Section~\ref{sec:methods}. Direct cross-model comparisons should account for potential feature set confounds.

\begin{table}[ht]
    \centering
    \caption{Best Tuned Model Results}
    \label{tab:results}
    \begin{tabular}{|l|l|l|l|l|l|l|}
        \hline
         Model & Accuracy & Precision & Recall & F1 & AUC-ROC \\
         \hline
         Rand Forest & 0.9934 & 0.9423 & 0.9612 & 0.9516 & 0.9937 \\
         GRU           & 0.9833 & 0.8938 & 0.8545 & 0.8737 & 0.9443 \\
         LSTM          & 0.9823 & 0.8704 & 0.8668 & 0.8686 & 0.9773 \\
         RNN           & 0.9829 & 0.9359 & 0.8018 & 0.8637 & 0.9110 \\
         Autoencoder   & 0.9181 & 0.4452 & 0.8427 & 0.5826 & 0.9050 \\
        \hline
    \end{tabular}
\end{table}

The Random Forest achieves the highest F1-Score (0.9516) and near perfect AUC-ROC (0.9937), benefiting from direct access to labeled training data. Among the unsupervised temporal models, the GRU attains the best F1-Score (0.8737), closely followed by the LSTM (0.8686) and RNN (0.8637). The feedforward Autoencoder achieves the lowest F1-Score (0.5826) but the fastest inference latency (0.039~ms). All temporal models maintain recall above 0.80, indicating robust detection of attack traffic despite the significant class imbalance. The improvement in recall for the recurrent models over the Autoencoder supports the hypothesis that sequential GOOSE message modeling captures temporal attack signatures that static, per-frame reconstruction cannot.

\begin{figure*}[h!]
    \centering
    \includegraphics[width=0.9\textwidth]{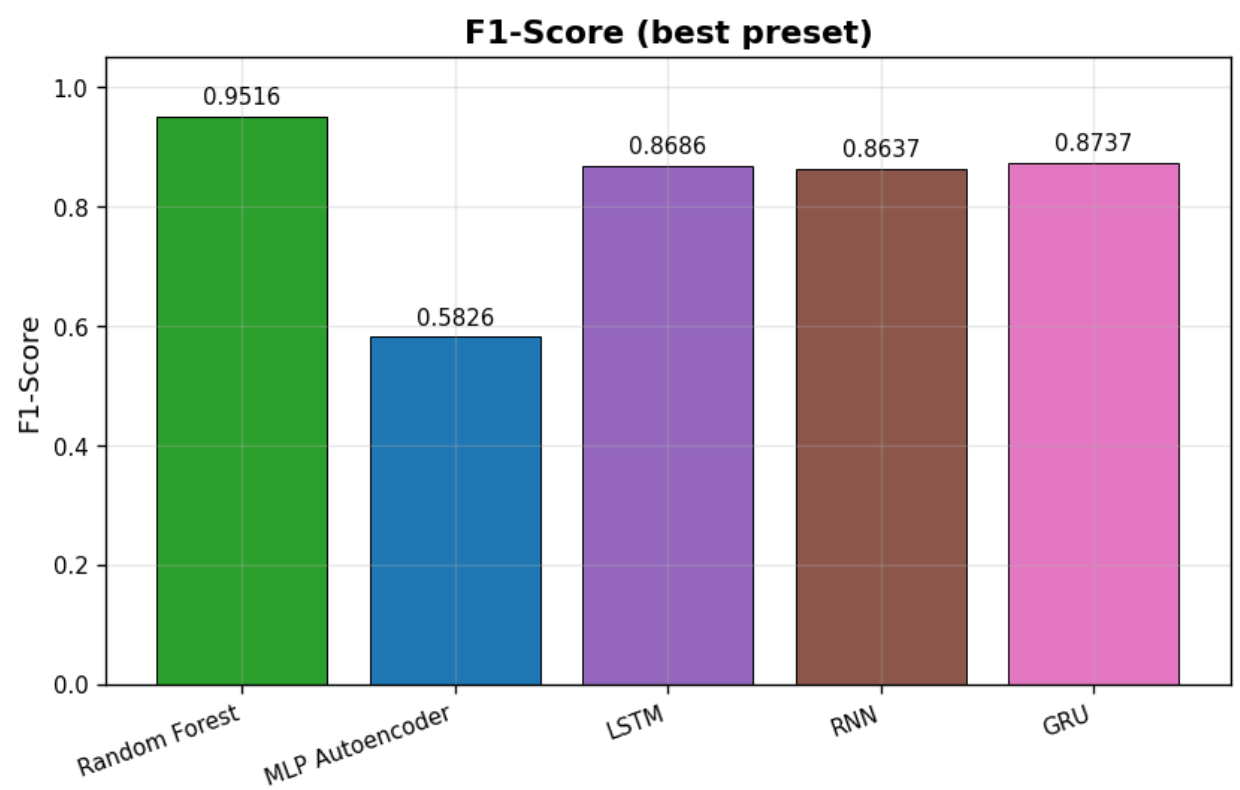}
    \caption{Model F1-Scores. The GRU leads among unsupervised models. The Random Forest is included for reference but does not satisfy the 4~ms latency requirement.}
    \label{fig:f1scores}
\end{figure*}

Figure~\ref{fig:f1scores} provides a visual comparison of F1-scores across all models.

\subsection{Inference Throughput and Latency}
Table~\ref{tab:throughput} reports single-sample inference latency in milliseconds for each model. The 4~ms GOOSE latency requirement is used as the deployment feasibility threshold. The feedforward Autoencoder is the fastest at 0.039~ms, followed by the GRU (1.118~ms), RNN (1.229~ms), and LSTM (1.920~ms). All four unsupervised models comfortably satisfy the 4~ms constraint. The Random Forest requires 21.837~ms per prediction and fails the real-time requirement, making it unsuitable for inline deployment without hardware acceleration or batching optimization.

\begin{table}[ht]
    \centering
    \caption{Model Inference Latency}
    \label{tab:throughput}
    \begin{tabular}{|l|l|l|}
        \hline
         Model & Latency (ms) & Meets $<$4 ms Requirement \\
         \hline
         Autoencoder   & 0.039  & \checkmark \\
         GRU           & 1.118  & \checkmark \\
         RNN           & 1.229  & \checkmark \\
         LSTM          & 1.920  & \checkmark \\
         Random Forest & 21.837 & $\times$   \\
        \hline
    \end{tabular}
\end{table}

\begin{figure*}[h!]
    \centering
    \includegraphics[width=0.9\textwidth]{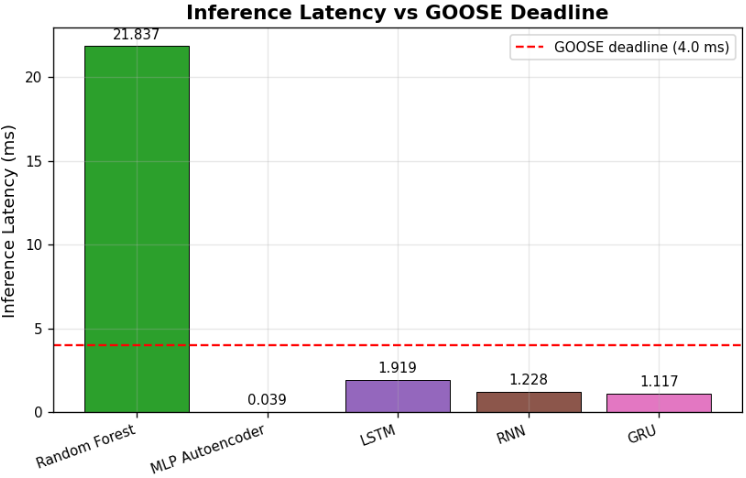}
    \caption{Model Latency. The dashed red line marks the 4~ms GOOSE protection threshold. All unsupervised models fall well below this limit. The Random Forest exceeds it by more than a factor of 5 times the threshold.}
    \label{fig:latency}
\end{figure*}

Latency measurements are hardware dependent. Results were obtained on a GPU-accelerated environment (Nvidia GeForce RTX 4070 Ti, 12~GB; Intel Core i7). Deployments on embedded substation hardware without GPU support should expect higher latency, particularly for recurrent models, and a hardware-specific evaluation is recommended before production deployment. The Random Forest latency is measured using scikit-learn's \texttt{predict} with batch size 1, which incurs non-trivial per-call overhead. Inference on batched or hardware optimized implementations would be faster, though the model would likely still not meet the 4~ms threshold without significant engineering effort.

\section{Discussion} \label{sec:discuss}

\begin{figure*}[h!]
    \centering
    \includegraphics[width=1.0\textwidth]{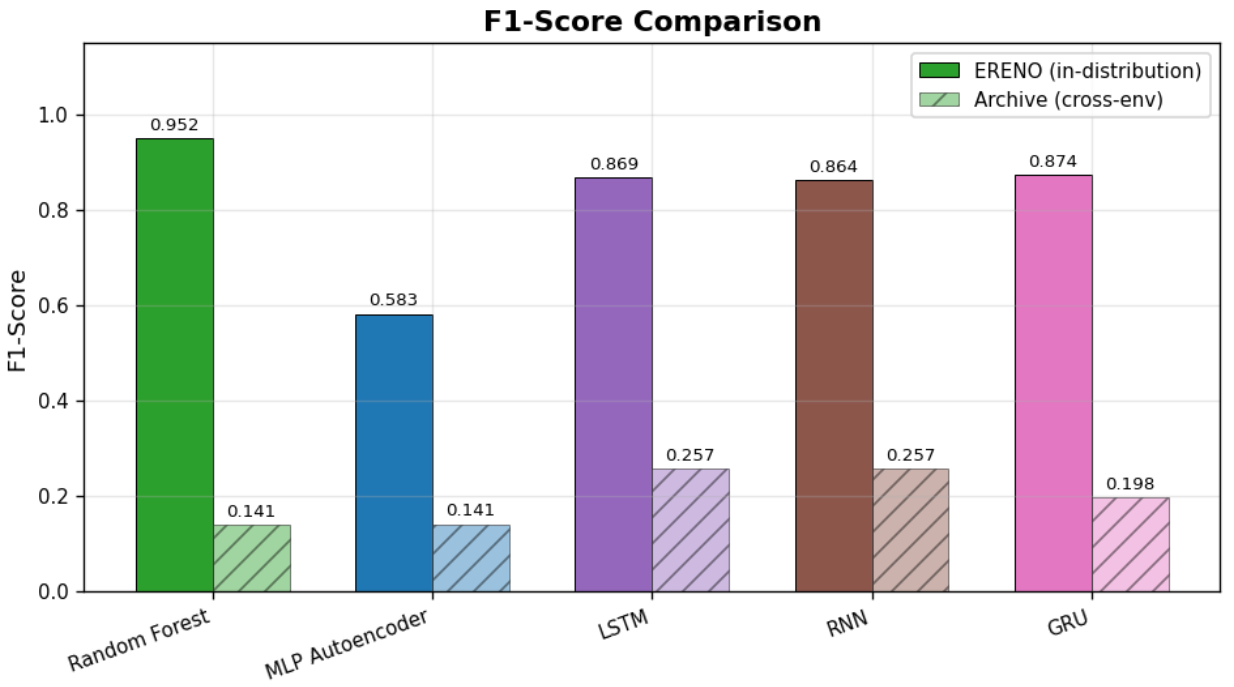}
    \caption{Model Generalizability. F1-score on the ERENO in-distribution test set (left) versus the de Oliveira cross-environment dataset (right). The Random Forest suffers the largest absolute drop. Recurrent models retain comparatively higher cross-environment performance.}
    \label{fig:generalizability}
\end{figure*}

\subsection{Effect of Sequence Window Length}
Hyperparameter tuning selected $SEQ\_LEN = 10$ as optimal for both the LSTM and RNN, while the GRU performed best at $SEQ\_LEN = 5$. This is consistent with the observation by de Oliveira et al.\ \cite{DEOLIVEIRA2025104197} that window length sensitivity varies across recurrent architectures. A window of 10 messages captures approximately one full GOOSE retransmission burst, providing sufficient temporal context to detect anomalous $StNum$ and $SqNum$ progressions without accumulating excessive inference overhead. Longer windows increased latency without a commensurate gain in detection performance under these experimental conditions.

\subsection{Model Selection and Deployment Tradeoffs}
The GRU (F1\,=\,0.8737, latency\,=\,1.118~ms) represents the best overall model for inline substation deployment, achieving the highest F1-score among unsupervised models while satisfying the 4~ms latency constraint with a comfortable margin. The LSTM achieves a comparable F1-score (0.8686) at slightly higher latency (1.920~ms), while the RNN offers competitive precision (0.9359) at 1.229~ms but lower recall (0.8018). The Autoencoder (0.039~ms) is the most latency efficient option and may be appropriate when computational resources are highly constrained, though its substantially lower F1 (0.5826) reflects difficulty distinguishing low rate attacks from normal traffic on a per-frame basis.

The Random Forest outperforms all unsupervised models in detection accuracy but at a latency exceeding the GOOSE protection threshold by more than five times. It may remain useful as a forensic or post-hoc analysis tool, but is not suitable for inline real-time IDS without hardware acceleration.

\subsection{Class Imbalance and Threshold Sensitivity}
With approximately 93\% normal traffic, anomaly threshold selection is critical. A naive threshold could achieve high accuracy while detecting few attacks. The Youden Index criterion applied on the validation partition partially mitigates this by jointly maximizing TPR and minimizing FPR. Nevertheless, practitioners deploying these models in environments with different imbalance ratios should expect threshold sensitivity and should recalibrate thresholds using a representative sample of local normal traffic before deployment.

\subsection{Reconstruction Error Analysis}
Reconstruction error distributions for normal and attack samples provide diagnostic insight into both autoencoder based models. Both the feedforward Autoencoder and the LSTM achieve AUC-ROC values above 0.87, confirming that reconstruction error is a reliable anomaly signal across the evaluated attack classes. The degree of separation varies by attack type. Event driven attacks such as Replay and Masquerade, which produce characteristic $StNum$ and $SqNum$ anomalies across message sequences, are most distinguishable by the temporal models. Instantaneous injection attacks may be equally detectable by the static Autoencoder.

\section{Cross-Environment Generalizability} \label{sec:generalize}

To assess how well models trained on ERENO transfer to a different deployment environment, all five models were evaluated without retraining on a second IEC-61850 dataset collected by de Oliveira et al.~\cite{DEOLIVEIRA2025104197} using a separate Hardware-in-the-Loop testbed with different IEDs and substation configurations. The second dataset covers four attack types: Masquerade, Message Injection, Poisoning, and Replay. Cross-environment evaluation is an important practical test because supervised models that overfit to dataset-specific feature distributions are known to degrade sharply when applied to new environments.

\subsection{Feature Coverage and Dataset Alignment}

Because the two datasets originate from different testbeds, not all ERENO features are present in the archive dataset. Feature coverage varies by preset: the standard GOOSE feature preset achieves 100\% coverage; the de Oliveira preset covers 87\% (13 of 15 features); the GOOSE+SV preset covers only 62\% (13 of 21 features), as the Sampled Values instantaneous features are absent from the archive dataset. Missing features were imputed with zeros.

This imputation introduces a potential confounding factor where performance degradation may partially reflect the shift in reconstruction error and decision boundary thresholds caused by filled in features, rather than purely reflecting distributional differences between testbeds. The most interpretable cross-environment comparison is therefore the GOOSE-only preset, which achieves full feature coverage. In a real world deployment, all protocol fields would be present by construction and per-environment threshold calibration on a small sample of normal traffic would be expected to recover a significant portion of the observed performance drop.

\subsection{Generalizability Results}

Table~\ref{tab:generalize} reports F1-Score and AUC-ROC for each model on the archive dataset alongside the corresponding ERENO in-distribution scores and the absolute F1-Score drop between environments. A visual comparison is provided in Figure~\ref{fig:generalizability}.

\begin{table}[ht]
    \centering
    \caption{Cross-Environment Generalizability (ERENO vs.\ de Oliveira Dataset)}
    \label{tab:generalize}
    \begin{tabular}{|l|l|l|l|}
        \hline
        Model & ERENO F1 & Oliveira F1 & F1 Drop \\
        \hline
        Rand Forest   & 0.9516 & 0.1414 & 0.8103 \\
        GRU             & 0.8737 & 0.1978 & 0.6759 \\
        LSTM            & 0.8686 & 0.2568 & 0.6118 \\
        RNN             & 0.8637 & 0.2568 & 0.6069 \\
        Autoencoder   & 0.5826 & 0.1414 & 0.4412 \\
        \hline
    \end{tabular}
\end{table}

All models experience substantial F1-Score degradation on the archive dataset without retraining, which is expected given differences in testbed configuration, traffic patterns, and feature availability. The Random Forest suffers the largest absolute drop (0.8103), collapsing from an F1-Score of 0.9516 to 0.1414. This is consistent with a supervised model's reliance on dataset specific statistical patterns. Features that are highly discriminative in ERENO do not necessarily carry the same signal in a different substation environment.

The recurrent models (LSTM and RNN, F1 drop $\approx 0.61$) retain higher cross-environment performance than the Random Forest, a result that holds even accounting for the zero-imputation confound affecting all models. This directional advantage suggests that learning over temporal sequences of GOOSE message behavior generalizes better than fitting static per-sample feature correlations to labeled attack classes. However, the magnitude of recurrent model generalizability should be interpreted cautiously until validated on a dataset with full feature overlap.

\section{Conclusion} \label{sec:conclusion}

This paper evaluated whether unsupervised temporal modeling offers practical advantages for intrusion detection in IEC-61850 GOOSE networks, with explicit attention to real-time latency constraints and cross-environment generalizability. Five models were implemented and evaluated on the ERENO IEC-61850 dataset: a supervised Random Forest baseline, a feedforward Autoencoder, and three recurrent sequence models (LSTM, RNN, and GRU). Anomaly thresholds for all unsupervised models were selected on a held-out validation partition to avoid test-set leakage.

All four unsupervised models satisfy the 4~ms GOOSE latency requirement, while the supervised Random Forest (despite achieving the highest F1-Score of 0.9516) requires 21.837~ms per prediction and is unsuitable for inline deployment. Among the unsupervised models, the GRU achieves the best F1-Score (0.8737) at 1.118~ms, followed closely by the LSTM (F1\,=\,0.8686, 1.920~ms) and RNN (F1\,=\,0.8637, 1.229~ms). The feedforward Autoencoder provides the fastest inference (0.039~ms) at lower detection performance (F1\,=\,0.5826), making it appropriate for the most resource constrained environments.

Cross-environment evaluation revealed that the Random Forest suffers the largest performance degradation when applied without retraining (F1 drop of 0.81), while recurrent temporal models retain comparatively higher cross-environment performance. This finding is directionally consistent with the expectation that training on normal traffic patterns generalizes better than fitting labeled attack distributions, though the magnitude of the advantage should be validated on datasets with full feature overlap. Based on the combined evidence, a GRU or RNN based unsupervised autoencoder is the recommended architecture for real-time anomaly detection in IEC-61850 GOOSE networks, with the GRU preferred when accuracy is the priority and the RNN when precision-oriented detection is needed.

Based on the results presented in this study, the GRU model appears best suited for real-time GOOSE intrusion detection, offering comparatively high detection accuracy and cross-environment generalizability while maintaining inference latency well within the 4~ms threshold required by the standard.

Future work could extend this evaluation to multi-class attack classification, investigate bidirectional LSTM and GRU variants as explored by de Oliveira et al.~\cite{DEOLIVEIRA2025104197}, and incorporate the explainability framework proposed by Lozano-Paredes et al.~\cite{lozano2026explainable} to provide interpretable alarms to substation operators. Validation on Hardware-in-the-Loop environments~\cite{ERENO2024,DEOLIVEIRA2025104197} and testing on a greater variety of substation configurations remain important steps toward production deployment confidence. Future work should also evaluate all model architectures on a single standardized feature set to isolate architectural contributions from feature set effects observed in the current study.

\printbibliography

\end{document}